\documentclass[12pt]{article} 

\usepackage[utf8]{inputenc}
\usepackage[T1]{fontenc} 
\usepackage{lmodern}
\usepackage[english]{babel}           

\usepackage{cancel}
\usepackage[pdftex]{graphicx}
\usepackage{epsfig}
\usepackage{graphicx}
\usepackage{comment}
\usepackage{latexsym}
\usepackage{hyperref}
\usepackage{amsmath}
\allowdisplaybreaks 
\usepackage{mathrsfs}
\usepackage[usenames, dvipsnames]{color}
\usepackage{amsbsy}
\usepackage{amssymb}
\usepackage{amsthm}
\usepackage{amsfonts}
\usepackage{cite}
\usepackage{enumitem}
\usepackage{xcolor}
\usepackage{color}
\usepackage{mathtools}
\usepackage{caption} 
\usepackage{subcaption} 
\usepackage{euscript} 

\usepackage{float}

\usepackage{diagbox}

\usepackage[title]{appendix}

\usepackage{tabularx}

\newcommand{\be}[0]{\begin{equation}}
\newcommand{\ee}[0]{\end{equation}}
\newcommand{\dis}{\displaystyle}
\renewcommand{\thefootnote}{\fnsymbol{footnote}}

\newcommand{\X}{Z}

\newcommand{\R}{\mathbb{R}}
\newcommand{\Z}{\mathbb{Z}}

\renewcommand{\O}{{\cal O}}

\renewcommand{\Im}{{\rm Im}\,}

\newcommand{\tr}{\textrm{tr}}

\newcommand{\rmv}{{\rm v}}

\newcommand{\via}{{\it via} }

\newcommand{\where}{\mbox{where}}

\renewcommand{\and}{\mbox{and}}

\newcommand{\esps}{\phantom{\!\!\!\overset{|}{a}}}
\newcommand{\esp}{\phantom{\!\!\overset{\displaystyle |}{|}}}

\newcommand{\bm}{\boldmath} 

\newcommand{\F}{{\cal F}}
\newcommand{\N}{{\cal N}}

\newcommand{\A}{{\cal A}}

\newcommand{\C}{{\cal C}}
\newcommand{\M}{{\cal M}}

\newcommand{\tl}{{\tilde l}}

\newcommand{\taudc}{\tau^{\rm dc}}

\newcommand{\nui}{\nu_{\rm int}}
\newcommand{\nue}{\nu_{\rm ext}}

\newcommand{\half}{\frac{1}{2}}

\newcommand\dd{\text{d}}

\def\jac(#1,#2){%
\begin{bsmallmatrix}
#1\cr 
#2\cr
\end{bsmallmatrix}}

\catcode`\@=11
\def\marginnote#1{}
\newcount\hour
\newcount\minute
\newtoks\amorpm
\hour=\time\divide\hour by60 \minute=\time{\multiply\hour by60
\global\advance\minute by-\hour}
\edef\standardtime{{\ifnum\hour<12 \global\amorpm={am}%
        \else\global\amorpm={pm}\advance\hour by-12 \fi
        \ifnum\hour=0 \hour=12 \fi
        \number\hour:\ifnum\minute<10 0\fi\number\minute\the\amorpm}}
\edef\militarytime{\number\hour:\ifnum\minute<10 0\fi\number\minute}
\def\draftlabel#1{{\@bsphack\if@filesw {\let\thepage\relax
   \xdef\@gtempa{\write\@auxout{\string
      \newlabel{#1}{{\@currentlabel}{\thepage}}}}}\@gtempa
   \if@nobreak \ifvmode\nobreak\fi\fi\fi\@esphack}
        \gdef\@eqnlabel{#1}}
\def\@eqnlabel{}
\def\@vacuum{}
\def\draftmarginnote#1{\marginpar{\raggedright\scriptsize\tt#1}}
\def\draft{\oddsidemargin -.2truein
        \def\@oddfoot{\sl preliminary draft \hfil
        \rm\thepage\hfil\sl\today\quad\militarytime}
        \let\@evenfoot\@oddfoot \overfullrule 3pt
        \let\label=\draftlabel
        \let\marginnote=\draftmarginnote
   \def\@eqnnum{(\theequation)\rlap{\kern\marginparsep\tt\@eqnlabel}%
\global\let\@eqnlabel\@vacuum}  }
\def\thebibliography#1{
\vskip 0.5cm \centerline{\bf \Large References}
\list{
[\arabic{enumi}]}{\settowidth\labelwidth{[#1]}
\leftmargin\labelwidth
\advance\leftmargin\labelsep
\usecounter{enumi}}
\def\newblock{\hskip .11em plus .33em minus .07em}
\sloppy\clubpenalty4000\widowpenalty4000
\sfcode`\.=1000\relax}

\renewcommand{\theequation}{\arabic{section}.\arabic{equation}}
\renewcommand{\section}{\setcounter{equation}{0}\@startsection
{section}{1}{0mm}{-\baselineskip}{0.5\baselineskip} {\normalfont\Large\bfseries}}
\renewcommand{\subsection}{\@startsection
{subsection}{2}{0mm}{-\baselineskip}{0.5\baselineskip} {\normalfont\large\bfseries}}
\renewcommand{\subsubsection}{\@startsection
{subsubsection}{3}{0mm}{-\baselineskip}{0.5\baselineskip}
{\normalfont\normalsize\slshape}}

\newcommand{\overbar}[1]{\mkern 1.5mu\overline{\mkern-1.5mu#1\mkern-1.5mu}\mkern 1.5mu}

\newcommand{\Xcl}{Z_{\text{cl}}}
\newcommand{\Xclbar}{\overbar{Z}_{\text{cl}}}
\newcommand{\Xqu}{Z_{\text{qu}}}
\newcommand{\Xqubar}{\overbar{Z}_{\text{qu}}}

\newcommand{\Scl}{S_{\text{cl}}}

\newcommand{\zu}{z_{1}}
\newcommand{\zd}{z_{2}}

\newcommand{\zud}{z_{12}}

\newcommand{\thetau}{\vartheta_{1}}
\newcommand{\thetauzud}{\vartheta_{1}(\zud)}

\makeatletter
\newcommand{\vast}{\bBigg@{3.5}}
\makeatother

\begin{document}


\begin{titlepage}
\begin{flushright}
December 2020
\vspace{1.5cm}
\end{flushright}
\begin{centering}
{\bm\bf \Large Two-point functions of Neumann--Dirichlet open-string \\  
\vspace{0.2cm}sector moduli\footnote{Based on a talk given at the “9th International Conference on New Frontiers in Physics'' (ICNFP 2020), 4–12 September 2020, Kolympari, Greece.}}

\vspace{7mm}

 {\bf Thibaut Coudarchet and Herv\'e Partouche}

 \vspace{4mm}

{CPHT, CNRS, Ecole Polytechnique, IP Paris, \\F-91128 Palaiseau, France \\ \vspace{2mm}
\textit{thibaut.coudarchet@polytechnique.edu}\\  \textit{herve.partouche@polytechnique.edu}}

\end{centering}
\vspace{0.1cm}
$~$\\
\centerline{\bf\Large Abstract}\\
\vspace{-1cm}

\begin{quote}

\hspace{.6cm} 

We compute at one loop the two-point functions of massless scalars in the Neumann--Dirichlet open-string sector of the type IIB orientifold compactified on $T^2\times T^4/\Z_2$, when $\N=2$ supersymmetry is spontaneously broken. This is done by evaluating correlation functions of ``boundary-changing vertex operators'' which are analogous to correlators of twist fields for closed strings. We use our results to compute the mass developed at one loop by the moduli fields arising in the Neumann--Dirichlet sector. 


\end{quote}

\end{titlepage}
\newpage
\setcounter{footnote}{0}
\renewcommand{\thefootnote}{\arabic{footnote}}
 \setlength{\baselineskip}{.7cm} \setlength{\parskip}{.2cm}

\setcounter{section}{0}


\section{Introduction}	

Vertex operators of open-string states in  Neumann--Dirichlet (ND) sectors change the nature of the boundary conditions at the insertion points. They  involve ``boundary-changing fields" whose operator-product expansions (OPE's) are analogous to those of twist fields\cite{Hashimoto}  which create states in closed-string twisted sectors of  orbifold theories\cite{Dixon}. Hence, computational techniques for twist fields\cite{Atick} can  be exploited, along with the method of images which relates correlation functions on open-string worldsheets to those for closed strings\cite{Burgess1,Burgess2}, to calculate string amplitudes involving states in ND sectors.

In this work, we are interested in the evaluation at one loop of the two-point functions of the massless scalars sitting in the ND sector of a type~IIB orientifold model compactified on $T^2\times T^4/\Z_2$,\cite{BianchiSagnotti, GimonPolchinski,GimonPolchinski2} where  $\N=2$ supersymmetry in Minkowski spacetime is spontaneously broken to $\N=0$ \via a string version\cite{openSS2,openSS3} of the Scherk-Schwarz mechanism\cite{SS1}. The computation of this amplitude  originates from the need to evaluate the masses generated at the quantum level by all moduli fields present in the model, in order  to perform a stability analysis at one loop\cite{ACP,CP,Coudarchet:2020ozn}. Since the dependance of the quantum effective potential on the vacuum expectation values of the moduli fields arising from the ND sector is unknown, the derivation of the quantum mass of these scalars  relies on a direct  computation of a two-point function.

In Sect.~\ref{model-amplitude}, we briefly review the relevant content and specificities of the \mbox{$\N=2\to\N=0$} model we are working in. We also  define the vertex operators of the massless scalars in the ND sector and derive the expression of their two-point functions in terms of various correlators. Sect.~\ref{twist-fields} paves the road to the evaluation of the correlators of boundary-changing field and Sect.~\ref{full-amplitude} makes use of these results to evaluate the two-point function we are interested in. In Sect.~\ref{limits}, we consider the regime of small supersymmetry breaking scale to obtain a more practical expression of the squared mass developed by the massless scalars of the ND sector. Further details beyond the results presented here can be found in Ref.~\cite{CP}.

\section{Two-point function of massless states\!\;in\!\;the \mbox{ND\!\;sector}}
\label{model-amplitude}

\subsection{The open-string model}

Our starting point is the type IIB orientifold model initially constructed in six dimensions by Bianchi and Sagnotti, and by Gimon and Polchinski\cite{BianchiSagnotti,GimonPolchinski,GimonPolchinski2}. Compactifying  down to four dimensions, the full $\N=2$ supersymmetric background is 
\begin{equation}
\R^{1,3}\times T^2\times\frac{T^4}{\Z_2}\, .
\end{equation}
The spacetime coordinates are labeled by Greek indices $\mu\in\{0,\dots,3\}$, while  the $T^2$ directions are denoted by primed Latin indices $I'\in\{4,5\}$. Moreover, the coordinates of $T^4$ are labeled by unprimed Latin indices $I\in\{6,7,8,9\}$ and modded by the \mbox{$\Z_2$-orbifold} generator
\begin{equation}
g:\quad (X^{6},X^{7},X^{8},X^{9})\longrightarrow(-X^{6},-X^{7},-X^{8},-X^{9})\, .
\end{equation}
Due to the orientifold projection, one O9-plane fills the entire space which requires the existence of 32 D9-branes for the model to be consistent. The orbifold action also implies the presence of one O5-plane at each fixed point of $T^4/\Z_2$, along with 32 D5-branes orthogonal to the $T^4$ directions. Eventually, $\N=2$ supersymmetry is spontaneously broken to $\N=0$ thanks to the  Scherk-Schwarz mechanism\cite{openSS2,openSS3} implemented along the coordinate $X^5$ of the two-torus. For simplicity throughout this work, we will consider the tori $T^2$ and $T^4$ to be products of circles whose radii are denoted $R_{I'}$ and $R_I$. In that case, the scale of supersymmetry breaking, which is given by the gravitini masses, takes the form 
\begin{equation}
M_{3/2}=\frac{1}{2R_5}\, .
\end{equation}
Indeed, this expression is identical to that encountered in field theory\cite{SS1} in $4+1$ dimensions, when superpartner bosonic and fermionic  fields are imposed periodic and antiperiodic boundary conditions along the circle $S^1(R_5)$, which implies a Kaluza--Klein mass gap. 

This model contains various kinds of moduli fields originating from the NN, DD and ND open-string sectors, as well as from the untwisted and twisted closed-string sectors\cite{ACP}. In the present work, we focus on those arising from the ND sector by considering all classically massless scalars realized as strings stretched between D9-branes and D5-branes\cite{CP}. We will compute their squared masses in backgrounds corresponding to extrema of the one-loop effective potential with respect to all moduli fields except $M_{3/2}$ in general.\footnote{$M_{3/2}$ undergoes a runaway behavior at one loop, unless the background exhibits a Bose-Fermi degeneracy at the massless level\cite{ACP,CP,Abel:2015oxa,SNS1}.} To characterize these backgrounds, it is convenient to consider two T-dual descriptions in which the moduli fields in the NN and DD sectors can be interpreted geometrically. 

T-dualizing $T^2$, the internal space takes the form \mbox{$\tilde T^2\times T^4/\Z_2$} modded by the involution $(\tilde X^{I'},X^{I})\to (-\tilde X^{I'},-X^{I})$ which has $2^6=64$ fixed points, where $\tilde T^2$ is the T-dual torus of coordinates $\tilde X^{I'}$\cite{review-3}. In this picture, there is one O3-plane at each fixed point and the 32 D5-branes become 32 D3-branes transverse to $\tilde T^2\times T^4/\Z_2$. Their positions in this internal space parametrize the expectation values of all the moduli fields arising from the DD sector of the initial theory. Similarly, denoting $\tilde T^4$ the four-torus of coordinates $\tilde X^I$ T-dual to $T^4$, one obtains by T-dualizing all six internal directions a description with internal space  $\tilde T^2\times \tilde T^4/\Z_2$ modded by $(\tilde X^{I'},\tilde X^{I})\to (-\tilde X^{I'},-\tilde X^{I})$. There is one O3-plane at each of the 64 fixed points and 32 D3-branes T-dual to the initial D9-branes. Their positions  in $\tilde T^2\times \tilde T^4/\Z_2$ parametrize all NN-sector moduli expectation values of the original description. The backgrounds we are interested in correspond to distributing all D3-branes \mbox{T-dual} to the D5-branes or D9-branes on the fixed points of their respective internal spaces. In that case, the gauge symmetries supported by the stacks of D3-branes are unitary. 
Labelling the 64 fixed points of each T-dual picture by two indices $ii'$, where \mbox{$i\in\{1,\dots,16\}$} for those of $T^4/\Z_2$ or $\tilde T^4/\Z_2$ and $i'\in\{1,2,3,4\}$ for those associated with the $\tilde T^2$ directions, the  brane configurations are characterized  by the numbers $2n_{ii'}$ and $2d_{ii'}$ of D3-branes T-dual to D9- and D5-branes located at the fixed points~$ii'$. In the following, we compute the two-point functions of the massless scalars in the ND+DN sector, which are in the bifundamental representation  of some $U(n_{i_0i_0'})\times U(d_{j_0i_0'})$ for some given $i_0,j_0$ and $i_0'$.\footnote{The two fixed points $i_0i_0'$ and $j_0i_0'$ have the same primed index for the number of massless scalars not to vanish\cite{ACP,CP}.}

\subsection{Vertex operators and amplitudes}

\begin{figure}[H]
\begin{center}
\raisebox{-0.5\height}{\includegraphics[scale=0.50]{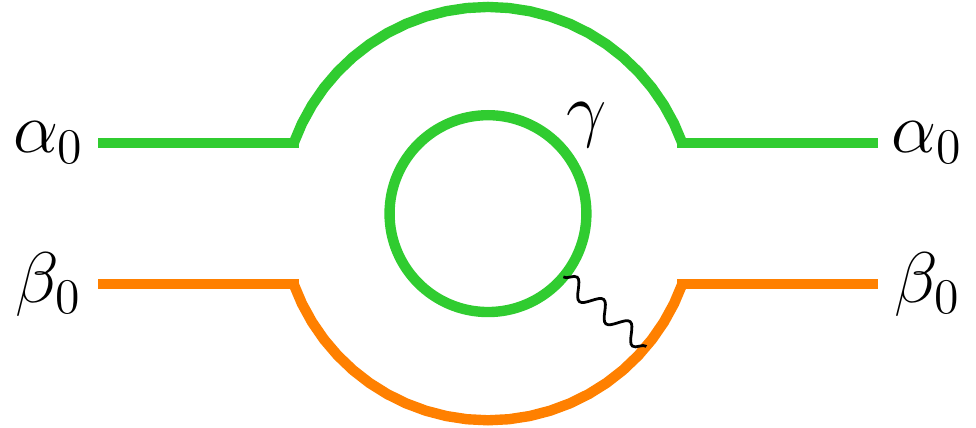}}
\quad~~~~~
\raisebox{-0.5\height}{\includegraphics[scale=0.50]{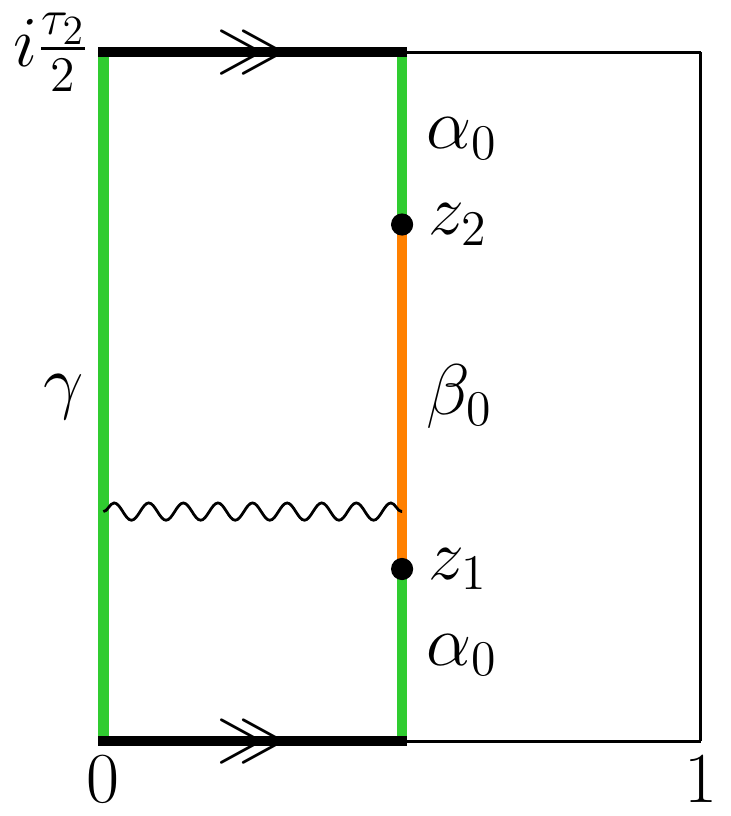}}\\[0.5cm]

\raisebox{-0.5\height}{\includegraphics[scale=0.50]{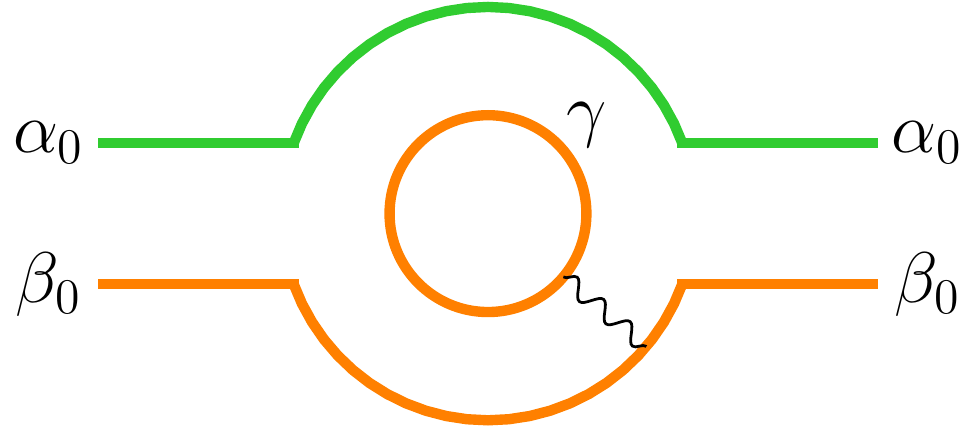}}
\quad~~~~~
\raisebox{-0.5\height}{\includegraphics[scale=0.50]{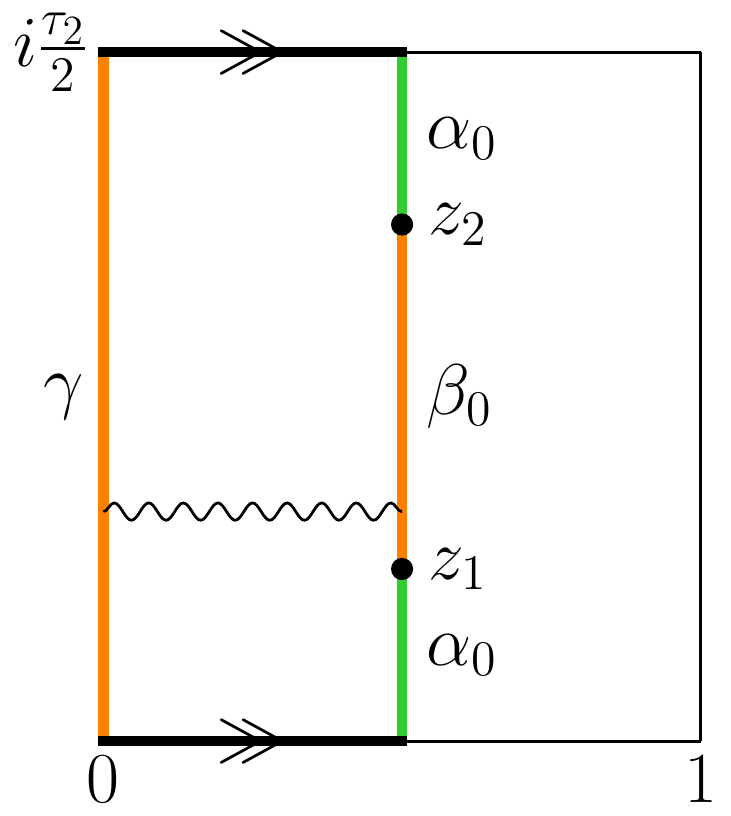}}\\[0.5cm]

\!\!\!\,\raisebox{-0.5\height}{\includegraphics[scale=0.50]{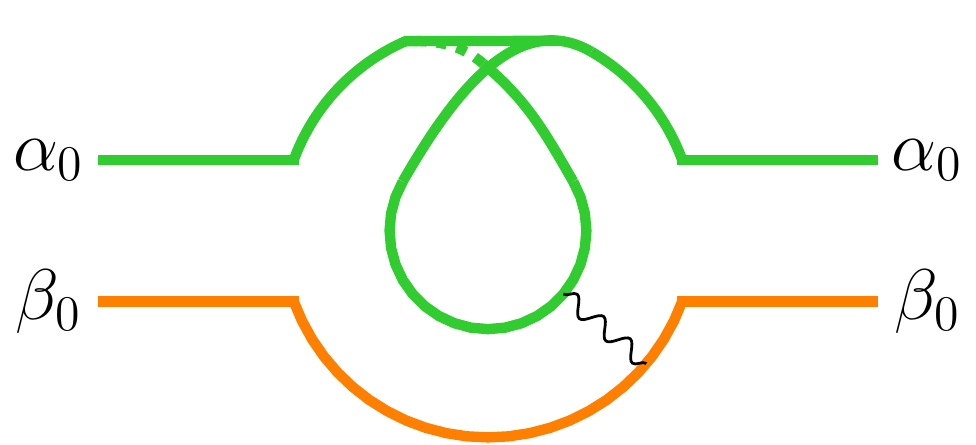}}
\!\!\!\!\!\!\!\!~~~\,
\raisebox{-0.5\height}{\includegraphics[scale=0.50]{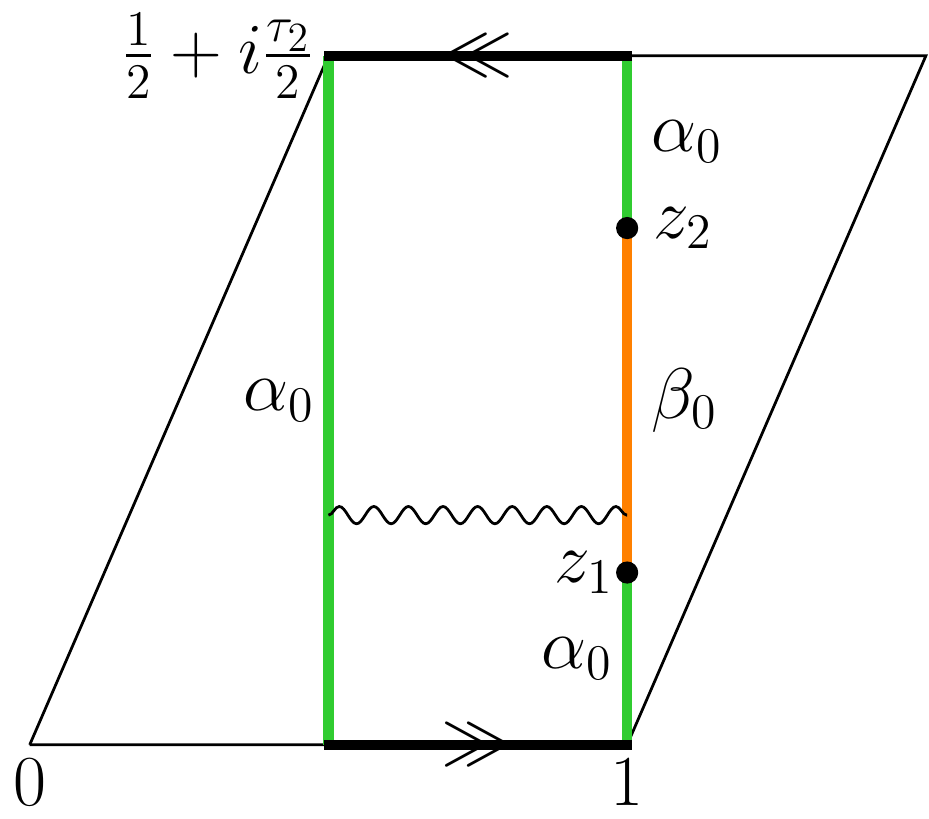}}
\end{center}
\caption{\footnotesize Open-string diagrams with two external legs in the ND and DN sectors (left panel). On the double-cover tori (right panel), the external legs are mapped to boundary-changing vertex operators at positions $z_1$, $z_2$.}
\label{annulus+mobius_diagrams}
\end{figure}

The amplitudes we consider, which  are depicted in the left panel of Fig.~\ref{annulus+mobius_diagrams}, are evaluated on worldsheets with topologies of the annulus or M\"obius strip.  After conformal transformations, the surfaces can be realized as double-cover tori with Teichm\"uller parameters 
\begin{equation}
\taudc=i{\tau_2\over 2} \mbox{ for the annulus} \quad \and\quad  \taudc=\half+i{\tau_2\over 2}\mbox{ for the M\"obius strip}\, ,\quad \tau_2>0\, ,
\end{equation}
modded by the involution $z\to 1-\bar z$, and with vertex operators inserted at $z_1, z_2$ (see the right panel of Fig.~\ref{annulus+mobius_diagrams}).  
The external legs bring a Chan-Paton index $\alpha_0$ referring to one of the $2n_{i_0i'_0}$ D9-branes (in green) and a Chan-Paton index $\beta_0$ which refers to one of the $2d_{j_0i'_0}$ D5-branes (in orange). For the annulus, the two legs must be attached to the same boundary, while the second boundary carries a Chan-Paton index $\gamma$ that refers either to one of the 32 D9-branes or one of the 32 D5-branes. There are therefore two annulus amplitudes to consider.

Along the boundaries where the vertices are located, the fact that the boundary condition changes from Neumann to Dirichlet and from Dirichlet to Neumann implies that the vertex operators involve boundary-changing fields. In ghost-picture~$-1$, the vertex operators actually take the form 
\begin{align}
\begin{split}
V_{-1}^{\alpha_0\beta_0}(\zu,k)&=\lambda_{\alpha_0\beta_0}\,e^{-\phi}\,e^{ik\cdot X}\, e^{{i\over2}(H_{3}-H_{4})}\, \sigma^{3}\sigma^{4}(\zu)\,,\\
V_{-1}^{\beta_0\alpha_0}(\zd,-k)&=\lambda^{\rm T}_{\beta_0\alpha_0}\,e^{-\phi}\,e^{-ik\cdot X}\,e^{-{i\over 2}(H_{3}-H_{4})}\, \sigma^{3}\sigma^{4}(\zd)\,,
\end{split}
\end{align}
where our definitions are as follows:
\begin{itemize}
\item $k^\mu$ is the external momentum satisfying on-shell the condition $k^\mu k_\mu=0$.

\item $\phi(z)$ is the ghost field encountered in the bosonization of the superconformal ghosts. 

\item $\lambda$ is the matrix that labels the states  transforming in the  bifundamental representation of $U(n_{i_0i'_0})\times  U(d_{j_0i'_0})$\cite{GimonPolchinski}, while $\lambda^{\rm T}$ stand for its transpose. 

\item Denoting $\psi^\mu(z)$, $\psi^{I'}(z)$, $\psi^I(z)$ the Grassmann fields superpartners of the bosonic  coordinates $X^\mu(z)$, $X^{I'}(z)$, $X^I(z)$, we define complex degrees of freedom for $u\in\{0,\dots,4\}$,
\begin{equation}
\begin{aligned}
&{\X}^u\equiv \frac{ X^{2u}+i X^{2u+1}}{\sqrt{2}}\, , &&\overbar{\X}^u\equiv \frac{ X^{2u}-iX^{2u+1}}{\sqrt{2}}\, ,\\
&{\Psi}^u\equiv  \frac{ \psi^{2u}+i\psi^{2u+1}}{\sqrt{2}}\equiv e^{iH_u}\, , &&\overbar{\Psi}^u\equiv \frac{\psi^{2u}-i\psi^{2u+1}}{\sqrt{2}}\equiv e^{-iH_u}\,,
\end{aligned}
\end{equation}
where $H_u$ are scalars introduced to bosonize the fermionic fields.

\item The operators $e^{\pm i( H_3-H_4)}$ are spin fields which enforce the fact that the ND- and DN-sector states we consider are spinors of the $T^4/\Z_2$ space.

\item $\sigma^u$, $u\in\{3,4\}$, are the boundary-changing fields associated with the complex directions~$Z^u$.
\end{itemize}

The computation of the amplitudes is more easily done with vertex operators in ghost-picture 0. In that case, the operators naturally split into sums of two contributions referred to as ``external'' and ``internal''. The former, $V_{0,\rm ext}^{\alpha_0\beta_0}(z_1,k), V_{0,\rm ext}^{\beta_0\alpha_0}(z_2,-k)$, involve the spacetime complexified momenta
\begin{equation}
K^u=\frac{k^{2u}+ik^{2u+1}}{\sqrt{2}}\, , \quad u\in\{0,1\}\, ,
\end{equation}
while the internal pieces, $V_{0,\rm int}^{\alpha_0\beta_0}(z_1,k), V_{0,\rm int}^{\beta_0\alpha_0}(z_2,-k)$, require the introduction of ``excited boundary-changing fields'' $\tau^u$ and $\tau'^u$, $u\in\{3,4\}$,  which appear in the OPE's
\begin{align}
\begin{split}
\label{OPE_dZ_sigma}
&\partial \X^u(z)\sigma^u(w)\underset{z\to w}{\sim} (z-w)^{-\half}\, \tau^u(w)+\mbox{finite}\, ,\\
&\partial\overbar{\X}^u(z)\sigma^u(w)\underset{z\to w}{\sim} (z-w)^{-\half}\,\tau^{\prime u}(w)+\mbox{finite}\,.
\end{split}
\end{align}
We thus define ``external'' and ``internal'' amplitudes to be evaluated on the annulus ($\Sigma=\A$) and M\"obius strip ($\Sigma=\M$). In terms of various correlators, their expressions  are given by 
\begin{align}
\begin{split}
\label{prelim_Aext}
A_{\text{ext}\Sigma}^{\alpha_0\beta_0}&\equiv \big\langle V_{0,\rm ext}^{\alpha_0\beta_0}(z_1,k)V_{0,\rm ext}^{\beta_0\alpha_0}(z_2,-k)\big\rangle^\Sigma\\
&=\alpha'\,\lambda_{\alpha_0\beta_0}\lambda^{\rm T}_{\beta_0\alpha_0}\,\langle e^{ik\cdot X}(\zu)e^{-ik\cdot X}(\zd)\rangle\,\langle e^{\frac{i}{2}H_3}(\zu)e^{-\frac{i}{2}H_3}(\zd)\rangle\langle e^{-\frac{i}{2}H_4}(\zu)e^{\frac{i}{2}H_4}(\zd)\rangle\times\esp\\
&\!\!\!\!\!\!\langle \sigma^{3}(\zu)\sigma^{3}(\zd)\rangle\,\langle \sigma^{4}(\zu)\sigma^{4}(\zd)\rangle\sum_{u=0}^1 |K^u|^2\Big[\langle e^{iH_{u}}(\zu)e^{-iH_{u}}(\zd)\rangle+\langle e^{-iH_{u}}(\zu)e^{iH_{u}}(\zd)\rangle\Big]\,,
\end{split}
\end{align}
and
\begin{align}
\begin{split}
\label{prelim_Aint}
A_{\text{int}\Sigma}^{\alpha_0\beta_0}&\equiv \big\langle V_{0,\rm int}^{\alpha_0\beta_0}(z_1,k)V_{0,\rm int}^{\beta_0\alpha_0}(z_2,-k)\big\rangle^\Sigma\\
&=\frac{1}{\alpha'}\,\lambda_{\alpha_0\beta_0}\lambda^{\rm T}_{\beta_0\alpha_0}\,\langle e^{ik\cdot X}(\zu)e^{-ik\cdot X}(\zd)\rangle\times \esp\\
&\,\;\;\;\;\;\;\Big[\langle e^{-\frac{i}{2}H_3}(\zu)e^{\frac{i}{2}H_3}(\zd)\rangle\,\langle e^{-\frac{i}{2}H_4}(\zu)e^{\frac{i}{2}H_4}(\zd)\rangle\,\langle \tau^{3}(\zu)\tau'^{3}(\zd)\rangle\,\langle \sigma^{4}(\zu)\sigma^{4}(\zd)\rangle\\
&\;\;\;\;+\langle e^{\frac{i}{2}H_3}(\zu)e^{-\frac{i}{2}H_3}(\zd)\rangle\,\langle e^{\frac{i}{2}H_4}(\zu)e^{-\frac{i}{2}H_4}(\zd)\rangle\,\langle \sigma^{3}(\zu)\sigma^{3}(\zd)\rangle\,\langle \tau'^{4}(\zu)\tau^{4}(\zd)\rangle\Big]\, .
\end{split}
\end{align}
In these formulas, sums over fermionic spin structures are implicit, while for $\Sigma=\A$ a sum over the second boundary $\gamma$ is also understood. Notice that it is the internal part of the amplitude, $A_{\text{int}\Sigma}^{\alpha_0\beta_0}$,  that captures the quantum mass we are looking for, while the external part, $A_{\text{ext}\Sigma}^{\alpha_0\beta_0}$, can be used to extract the one-loop correction to the K\"ahler potential of the scalars in the ND+DN sector.

\section{Twist-field correlators at genus-1}
\label{twist-fields}

The main difficulty in the evaluation of the open-string amplitudes (\ref{prelim_Aext}) and (\ref{prelim_Aint}) resides in the computation of the correlators of boundary-changing fields, which are related to their  counterparts for closed strings  we now focus on. 

\subsection{Ground-state twist fields}

The OPE's of $\partial \X^u(z)$ and $ \partial \overbar\X^u(z)$ with $\sigma^u(z)$, $u\in\{3,4\}$,  are identical to those with $\Z_2$-twist fields $\sigma^u(z,\bar z)$  which create the ground state of the closed-string twisted sector of a $T^4/\Z_2$ orbifold\cite{Hashimoto}. As a consequence, techniques to evaluate correlators of twist fields\cite{Dixon,Atick} can be adapted to  the case of boundary-changing fields.

The correlator of two $\Z_2$-twist fields on a genus-1 Riemann surface $\Sigma$ can be written as a sum over instanton contributions.
Decomposing the complex coordinates into a classical background and a quantum fluctuation, 
\begin{equation}
\label{cl+qu}
\X^u(z,\bar{z})=\Xcl^u(z,\bar{z})+ \Xqu^u(z,\bar{z})\,,
\end{equation}
we have 
\begin{equation}
\langle\sigma^u(\zu,\bar{z}_1)\sigma^u(\zd,\bar{z}_2)\rangle=\sum_{\Xcl^u}e^{-\Scl^{\Sigma u}}\langle\sigma^u(\zu,\bar{z}_1)\sigma^u(\zd,\bar{z}_2)\rangle_{\text{qu}}\, ,
\end{equation}
where $\Scl^{\Sigma u}$ is the classical action of $Z^u$, 
\begin{equation}
\label{classical_action}
S_{\text{cl}}^{\Sigma u}=\frac{i}{2\pi\alpha'}\int_{\Sigma}\dd z\wedge \dd\bar z\,\big(\partial\Xcl^u\, \bar \partial\Xclbar^u+\partial\Xclbar^u\,\bar \partial\Xcl^u\big)\,,
\end{equation}
and $\langle \sigma^u\sigma^u\rangle_{\rm qu}$ is the ``quantum correlator'' arising  from  quantum fluctuations. The latter have been computed in Ref.~\cite{Atick} for an arbitrary number of twist fields by using the stress-tensor method that we summarize briefly.  

The starting point is the ``Green's function in presence of twist fields'' \begin{equation}
g(z,w)\equiv\frac{\langle -\partial \X^u_{\rm qu}(z)\,\partial \overbar{\X}^u_{\rm qu}(w)\prod_A\sigma^u(z_A,\bar{z}_A)\rangle_{\rm qu}}{\alpha'\, \langle\prod_A\sigma^u(z_A,\bar{z}_A)\rangle_{\rm qu}}\,.
\label{gf}
\end{equation}
Upon subtracting the double-pole singularity as $z\to w$, this expression can be used to derive the quantity
\begin{equation}
\langle\langle T^u(z)\rangle\rangle\equiv\frac{\left\langle T^u(z)\prod_{A}\sigma^u(z_A,\bar{z}_A)\right\rangle_{\rm qu}}{\left\langle \prod_{A}\sigma^u(z_A,\bar{z}_A)\right\rangle_{\rm qu}}\,,
\end{equation}
thanks to the OPE of $\partial\Xqu(z)\partial\overbar{Z}_{\text{qu}}(w)$ which involves the stress tensor $T^u(w)$. Then, exploiting the OPE of $T^u(z)\sigma^u(z_A,\bar z_A)$ as $z$ approaches $z_A$, one finds differential equations
\begin{equation}
\partial_{z_B}\ln\big\langle\prod_A\sigma^u(z_A,\bar{z}_A)\big\rangle_{\rm qu}=\lim_{z\rightarrow z_B}\left[(z-z_B)\,\langle\langle T^u(z)\rangle\rangle-\frac{h}{(z-z_B)}\right],
\end{equation}
where $h=1/8$ is the conformal weight of the twist fields. Solving  these  equations, one obtains the desired correlators. The key point is that $g(z,w)$, which is needed for applying this program, can be determined by writing the most general function satisfying double periodicity on the torus $\Sigma$  and with correct behavior dictated by the OPE's as $z$ or $w$ approach some $z_A$. This can be done since functions with these properties form a vectorial space that can be expanded against a basis of so-called ``cut differentials,'' which are holomorphic one-forms on $\Sigma$. The coefficients of this expansion can be determined uniquely by imposing that $Z^u_{\rm qu}(z,\bar z)$ is doubly periodic. 

In our case of interest, one obtains\cite{Atick}
\begin{equation}
\langle \sigma^u(z_{1},\bar z_1)\sigma^u(z_{2},\bar z_2)\rangle_{\text{qu}}=f(\taudc)\, (\det W)^{-1}\, \thetau(z_{1}-z_{2})^{-\frac{1}{4}}\; \overbar{\thetau(z_{1}-z_{2})}^{-\frac{1}{4}}\, ,
\end{equation}
where $f(\taudc)$ is an ``integration constant'' that depends on the Teichm\"uller parameter $\taudc$ of $\Sigma$ and $\vartheta_\nu$, $\nu\in\{1,2,3,4\}$,  are the Jacobi modular forms. Moreover, $W$ is a $2\times 2$ matrix whose entries are the periods of the cut differential  
\begin{equation}
\omega(z)=\vartheta_{1}(z-z_{1})^{-\half}\, \vartheta_{1}(z-z_{2})^{-\half}\,\vartheta_{1}\big(z-\frac{z_{1}+z_{2}}{2}\big)
\end{equation} 
along the cycles $\gamma_{1}$:  $z\rightarrow z+1$ and $\gamma_{2}$:  $z\rightarrow z+\taudc$,
\begin{equation}
W=
\begin{pmatrix}
W_1 & \overbar{W}_1\\
W_2 & \overbar{W}_2
\end{pmatrix},\quad  \quad W_{a}=\oint_{\gamma_a}\dd z\, \omega\,,\quad a\in\{1,2\}\,.
\end{equation}

\subsection{Excited twist fields}

To compute correlators of excited twist fields, we use the OPE's (\ref{OPE_dZ_sigma}) to write\cite{Abel}
\begin{align}
\langle \tau^u(z_{1},\bar z_1)\tau^{\prime u}(z_{2},\bar z_2)\, \rangle_{\text{qu}}&=\lim_{\substack{z\rightarrow z_{1}\\ w\rightarrow z_{2}}}\!\left[(z-\zu)^{\half}(w-\zd)^{\half}\langle \partial Z^u(z)\partial \overbar{Z}^u(w)\sigma^u(z_{1},\bar z_1)\sigma^u(z_{2},\bar z_2)\rangle_{\rm qu}\right] .
\end{align}
Using the decomposition (\ref{cl+qu}), this expression splits naturally into two pieces 
\begin{align}
\begin{split}
\langle \tau^u(z_{1},\bar z_1)\tau^{\prime u}(z_{2},\bar z_2)\rangle_{\text{qu}}^{(1)}&=\langle \sigma^u(z_{1},\bar z_1)\sigma^u(z_{2},\bar z_2)\rangle_{\rm qu}\lim_{\substack{z\rightarrow z_{1}\\ w\rightarrow z_{2}}}\!\left[(z-\zu)^{\half}(w-\zd)^{\half}\,\partial \Xcl^u(z)\partial\Xclbar^u(w)\right] ,\\
\langle \tau^u(z_{1},\bar z_1)\tau^{\prime u}(z_{2},\bar z_2)\, \rangle_{\text{qu}}^{(2)}&=\lim_{\substack{z\rightarrow z_{1}\\ w\rightarrow z_{2}}}\!\left[(z-\zu)^{\half}(w-\zd)^{\half}\langle \partial \Xqu^u(z)\partial \Xqubar^u(w)\sigma^u(z_{1},\bar z_1)\sigma^u(z_{2},\bar z_2)\rangle_{\rm qu}\right] .
\end{split}
\end{align}

To compute part $(1)$ of the correlator, we determine $\partial\Xcl^u(z)$ and $\bar\partial\Xcl^u(\bar z)$ by using the fact that they must be linear sums of cut differentials whose coefficients imply that $Z^u_{\rm cl}(z,\bar z)$ are instanton solutions in the target space $S^1(R_{2u})\times S^1(R_{2u+1})$.  Denoting winding and wrapping numbers as $n_I, l_I\in\Z$, this conditions takes the form 
\begin{equation}
\oint_{\gamma_a}\dd z\, \partial\Xcl^u(z)+\oint_{\gamma_a}\dd\bar{z}\, \bar \partial\Xcl^u(\bar z)=v^u_{a}\,,\quad a\in\{1,2\}\, ,
\end{equation}
where we have defined 
\begin{equation}
v_{1}^{u}=\frac{2\pi R_{2u}n_{2u}+2i\pi R_{2u+1}n_{2u+1}}{\sqrt{2}}\, ,~~\quad v_{2}^{u}=\frac{2\pi R_{2u}l_{2u}+2i\pi R_{2u+1}l_{2u+1}}{\sqrt{2}}\, .
\end{equation}
Using these notations, we obtain that 
\begin{align}
\langle \tau^u(z_{1},\bar z_1)\tau^{\prime u}(z_{2},\bar z_2)\rangle_{\text{qu}}^{(1)}=i\,{(W^{-1})_{1}}^{a}v_a^u{(\overbar W^{-1})_{2}}^{b}\bar v_b^u&\,\frac{\thetau(\frac{\zu-\zd}{2})^{2}}{\thetau'(0)\, \thetau(\zu-\zd)}\langle \sigma^u(z_{1},\bar z_1)\sigma^u(z_{2},\bar z_2)\rangle_{\rm qu}\, .
\label{partie1}
\end{align}
Notice that the determination of $\partial\Xcl^u(z)$ and $\bar\partial\Xcl^u(\bar z)$ also yield an explicit expression of the action~(\ref{classical_action}).

Part $(2)$ of the correlator can be expressed in terms of the Green's function~(\ref{gf}). Using the explicit form of $g(z,w)$ given in Ref.~\cite{Atick}, one obtains
\begin{align}
\begin{split}
\langle \tau^u(z_{1},\bar z_1)\tau^{\prime u}(z_{2},\bar z_2)\, \rangle_{\text{qu}}^{(2)}=-i\,\alpha' &\langle \sigma^u(z_{1},\bar z_1)\sigma^u(z_{2},\bar z_2)\rangle_{\rm qu}\\
&\times\Bigg[C\frac{\thetau(\frac{z_1-z_2}{2})^{2}}{\thetau'(0)\,\thetau(z_1-z_2)}+\frac{\thetau'(0)\, F_{1}(\zu,\zd)}{2\, \thetau(z_1-z_2)}\Bigg]\, ,
\label{partie2}
\end{split}
\end{align}
where we have defined 
\begin{align}
C\,\frac{\thetau(\frac{\zu-\zd}{2})^{2}}{\thetau'(0)\,\thetau(\zu-\zd)} = -\half \, \vartheta_1'(0)\, \vartheta_1&\big({z_1-z_2\over 2}\big) {(W^{-1})_{1}}^{a}\oint_{\gamma_a}\dd z \,{F_1(z,z_2)\over \vartheta_1(z-z_1)^\half \, \vartheta_1(z-z_2)^{3\over 2}}\, .
\end{align}
In these expressions, the function $F_1(z,w)$ is given by 
\begin{equation}
F_1(z,w)=\frac{\vartheta_1\left(z-w\right)}{\vartheta_1\left(U_1\right)}\frac{\vartheta_1\left(z-w+\frac{\zd-\zu}{2}-U_1\right)}{\vartheta_1\left(\frac{\zd-\zu}{2}-U_1\right)}\, , 
\end{equation}
where $U_1$ is such that $\partial_{z}F_{1}(z,w)\big|_{z=w}=0$.

\section{Full open string amplitudes}
\label{full-amplitude}

The instanton action and correlators for a genus-1 worldsheet can be used to derive the open string amplitudes~(\ref{prelim_Aext}) and~(\ref{prelim_Aint}).

First of all, open-string instantons exist for worldsheets with NN, DD, ND or DN boundary conditions in the annulus case, and N or D boundary conditions for the M\"obius strip. In the  NN and N case, all winding numbers $n_{I'},n_I$ must vanish, while for DD or D boundary conditions, denoting with tildes the T-dual winding and wrapping numbers,  $\tilde n_{I'}$ and $\tilde n_I$ must also be zero. For ND and DN boundary conditions, instantons can only wrap $T^2$. Dividing by 2 the double-cover worldsheet classical action (\ref{classical_action}) for $u\in\{2,3,4\}$, one obtains for $\Sigma\in\{\A,\M\}$ the full open-string instanton actions
\begin{equation}
S^\Sigma_{\text{cl}}={\pi[(R_4l_4)^2+(R_5l_5)^2]\over \alpha' \tau_2}+{|W_1|^2\over 4\pi\alpha'\,{\text{Im}}(\overbar W_{1}W_{2})}\times \left\{ \begin{array}{ll}{\dis\sum_{u=3}^4 |v_{2}^u|^{2}} & \mbox{for NN and N ,}\\
{\dis\sum_{u=3}^4 |\tilde v_{2}^u|^{2}}& \mbox{for DD and D ,}\\
0 & \mbox{for ND and DN ,}\esps\end{array}
\right.
\end{equation}
where we have defined  for $u\in\{3,4\}$ 
\begin{equation}
 v_{2}^{u}=\frac{2\pi R_{2u}l_{2u}+2i\pi R_{2u+1}l_{2u+1}}{\sqrt{2}}\, ,\quad  \tilde v_{2}^{u}=\frac{2\pi { \alpha'\over R_{2u}}\, \tilde l_{2u}+2i\pi {\alpha'\over R_{2u+1}}\,\tilde l_{2u+1}}{\sqrt{2}}\, .
\end{equation}

Second, the method of images can be applied during the course of the derivation of the ground-state twist-field correlators to obtain those of ``ground-state boundary-changing fields.''  The result amounts to take the ``square root" of the closed-string result\cite{Abel}
\begin{equation}
\langle \sigma^u(z_{1})\sigma^u(z_{2})\rangle_{\text{qu}}=f_{\rm op}(\taudc)\,(\det W)^{-\half}\,  \thetau(z_{1}-z_{2})^{-\frac{1}{4}}\, ,
\end{equation}
where $f_{\rm op}(\taudc)$ depends on the Teichm\"uller parameter of the double-cover torus. The twist fields $\tau^u(z,\bar z), \tau^{\prime u}(z,\bar z)$ being excited only on their holomorphic sides, their correlators are identical to those of ``excited boundary-changing fields'', up to an extra division by 2 of part $(1)$ in Eq.~(\ref{partie1})  which follows from the same rescaling of the instanton action. 

Gathering all results and denoting  $\zud\equiv\zu-\zd$, we find the following expression for the external part of the amplitudes
\begin{align}
\begin{split}
A_{\text{ext}\Sigma}^{\alpha_0\beta_0}=&\; \alpha'k^2\, \lambda_{\alpha_0\beta_0}\lambda^{\rm T}_{\beta_0\alpha_0}\left(\left|\frac{\thetauzud}{\thetau'(0)}\right|e^{-\frac{2\pi}{\tau_2}[\text{Im}(\zud)]^{2}}    \right)^{\!-2\alpha'k^{2}}\, {1\over \det W\, \thetauzud^{2}}\times \\
& \sum_{\nue\neq 1}\, \vartheta_1'(0)\frac{\vartheta_{\nue}(z_{12})}{\vartheta_{\nue}(0)}\,\sum_{\nui}(-1)^{\delta_{\nui,1}}\, \vartheta_{\nui}\big({\zud\over 2}\big)^2\sum_{\vec l'}e^{-{\pi\over \alpha'\tau_2}\sum_{I'}(R_{I'}l_{I'})^2}\times\\
&\Bigg[\sum_{\vec{l}}e^{-\frac{|W_{1}|^{2}(|{v}_{2}^{3}|^{2}+|{v}_{2}^{4}|^{2})}{4\pi\alpha'\Im(\overbar{W}_1W_2)}}\C_{\nui}^{\Sigma\vec l'\vec l}+\sum_{\vec \tl}e^{-\frac{|W_{1}|^{2}(|\tilde{v}_{2}^{3}|^{2}+|\tilde{v}_{2}^{4}|^{2})}{4\pi\alpha'\Im(\overbar{W}_1W_2)}}\tilde \C_{\nui}^{\Sigma\vec l'\vec \tl}\Bigg]\, ,
\end{split}
\end{align}
while for  the internal parts we obtain
\begin{align}
\begin{split}
A_{\text{int}\Sigma}^{\alpha_0\beta_0}=&-\frac{s\,i}{\alpha'}\, \lambda_{\alpha_0\beta_0}\lambda^{\rm T}_{\beta_0\alpha_0}\left(\left|\frac{\thetauzud}{\thetau'(0)}\right|e^{-\frac{2\pi}{\tau_2}[\text{Im}(\zud)]^{2}}    \right)^{\!-2\alpha'k^{2}}\, {\vartheta_1({z_{12}\over 2})^2\over \det W\, \thetauzud^{2}\,  \vartheta_1'(0)} \times\\
& \; 4\,\sum_{\nui} \vartheta_{\nui}\big({\zud\over 2}\big)^2\, \sum_{\vec l'}e^{-{\pi\over \alpha'\tau_2}\sum_{I'}(R_{I'}l_{I'})^2}\times\esps \\
&\;\Bigg\{\:\sum_{\vec l}e^{-\frac{|W_{1}|^{2}(|{v}_{2}^{3}|^{2}+|{v}_{2}^{4}|^{2})}{4\pi\alpha'\Im(\overbar{W}_1W_2)}}\C_{\nui}^{\Sigma\vec l'\vec l}\Bigg[ {\overbar{W}_1^2\, (|v_2^3|^2+|v_2^4|^2)\over 8[\Im (\overbar{W}_1 W_2)]^2}+2\alpha'(C+\hat C)\Bigg]\\
&\;+\sum_{\vec \tl}e^{-\frac{|W_{1}|^{2}(|{\tilde v}_{2}^{3}|^{2}+|{\tilde v}_{2}^{4}|^{2})}{4\pi\alpha'\Im(\overbar{W}_1W_2)}}\tilde \C_{\nui}^{\Sigma\vec l'\vec \tl}\Bigg[{\overbar{W}_1^2\, (|\tilde v_2^3|^2+|\tilde v_2^4|^2)\over 8[\Im (\overbar{W}_1 W_2)]^2}+2\alpha'(C+\hat C)\Bigg]\Bigg\}\, ,
\label{Aint}
\end{split}
\end{align}
where 
\begin{equation}
\hat C\equiv {\thetau'(0)^2\over 2\,\vartheta_1({z_{12}\over 2})^2}\, F_{1}(\zu,\zd)\, .
\end{equation}
In these formulas, $\nue, \nui\,\in\{1,2,3,4\}$ label the spin structures of the worldsheet fermions $\Psi^0,\Psi^1,\Psi^2$ and $\Psi^3,\Psi^4$, respectively. Moreover, we have introduced normalization functions $\C_{\nui}^{\Sigma\vec l'\vec l}$ and $\tilde \C_{\nui}^{\Sigma\vec l'\vec \tl}$, which are the $\taudc$-dependent ``integration constants'' arising in the stress-tensor method applied for the correlators of the boundary-changing field and spin-fields. They can be identified by taking the limit $z_{12}\to 0$ in the external parts of the amplitudes which then reduce to sectors of the known one-loop partition function. In the annulus case, the results are 
\begin{align}
\label{Cnu}
&\C_{1}^{\A\vec l' \vec l}={\C\over \tau_2^2\, \eta^3}\, f_{\alpha_0\text{D}}^{\A\vec l' \vec l}\, ,&&\qquad\tilde{\C}_{1}^{\A\vec l'\vec \tl}={\C\over \tau_2^2\, \eta^3}\, f_{\beta_0\text{N}}^{\A\vec l'\vec \tl}\, ,\nonumber\\
&\C_{2}^{\A\vec l'\vec l}={\C\over \tau_2^2\, \eta^3}\,{\vartheta_3^2\over \vartheta_4^2}\,  f_{\alpha_0\text{D}}^{\A\vec l' \vec l}-{\C\,\vartheta_2^2\over \tau_2^4\, \eta^9}\, f_{\alpha_0\text{N}}^{\A\vec l'\vec l}\, e^{2i\pi \vec l'\cdot \vec a_{\rm S}'}\, ,&&\qquad\tilde{\C}_{2}^{\A\vec l'\vec \tl}={\C\over \tau_2^2\, \eta^3}\, {\vartheta_3^2\over \vartheta_4^2}\,f_{\beta_0\text{N}}^{\A\vec l'\vec \tl}-{\C\, \vartheta_2^2\over \tau_2^4\, \eta^9}\, f_{\beta_0\text{D}}^{\A\vec l'\vec \tl}\, e^{2i\pi \vec l'\cdot \vec a_{\rm S}'}\, ,\nonumber\\
&\C_{3}^{\A\vec l'\vec l}={\C\, \vartheta_3^2\over \tau_2^4\, \eta^9}\, f_{\alpha_0\text{N}}^{\A\vec l'\vec l}-{\C\over \tau_2^2\, \eta^3}\, {\vartheta_2^2\over \vartheta_4^2}\, f_{\alpha_0\text{D}}^{\A\vec l'\vec l}\, e^{2i\pi \vec l'\cdot \vec a_{\rm S}'}\, ,&&\qquad \tilde{\C}_{3}^{\A\vec l'\vec \tl}={\C\, \vartheta_3^2\over \tau_2^4\, \eta^9}\, f_{\beta_0\text{D}}^{\A\vec l'\vec \tl}-{\C\over \tau_2^2\, \eta^3}\, {\vartheta_2^2\over \vartheta_4^2}\, f_{\beta_0\text{N}}^{\A\vec l'\vec \tl}\, e^{2i\pi \vec l'\cdot \vec a_{\rm S}'}\, ,\nonumber \\
&\C_{4}^{\A\vec l'\vec l}=-{\C\, \vartheta_4^2\over \tau_2^4\, \eta^9}\, f_{\alpha_0\text{N}}^{\A\vec l'\vec l}\,,&&\qquad \tilde{\C}_{4}^{\A\vec l'\vec \tl}=-{\C\, \vartheta_4^2\over \tau_2^4\, \eta^9}\, f_{\beta_0\text{D}}^{\A\vec l'\vec \tl}\, ,
\end{align}
where $\C$ is a numerical factor, $\vec a_{\rm S}'$ is the two-vector whose  components are \mbox{$(a_{\rm S}^{\prime 4},a_{\rm S}^{\prime 5})=(0,\half)$} and 
\begin{align}
&f_{\alpha_0\text{N}}^{\A\vec l'\vec l}={\rmv_2\rmv_4\over \alpha'^3}\, \sum_{i,i'}2n_{ii'}\,e^{2i\pi\vec{l}\cdot(\vec{a}_{i_0}-\vec{a}_i)} e^{2i\pi\vec l'\cdot(\vec{a}_{i'_0}-\vec a_{i'})},&&f_{\beta_0\text{\text{D}}}^{\A\vec l'\vec \tl}={\rmv_2\,\alpha'^2\over \alpha'\,\rmv_4}\, \sum_{i,i'}2d_{ii'}\, e^{2i\pi\vec \tl\cdot(\vec{a}_{j_0}-\vec{a}_i)} e^{2i\pi\vec l'\cdot(\vec{a}_{i'_0}-\vec a_{i'})},\nonumber\\
&f_{\alpha_0\text{D}}^{\A\vec l'\vec{l}}=\delta_{\vec{l},\vec{0}}\, {\rmv_2\over \alpha'}\, \sum_{i,i'}2d_{ii'}\, e^{2i\pi \vec l'\cdot (\vec a_{i_0'}-\vec a_{i'})} ,&&f_{\beta_0\text{N}}^{\A\vec l'\vec \tl}=\delta_{\vec \tl,\vec{0}}\, {\rmv_2\over \alpha'}\, \sum_{i,i'}2n_{ii'}\, e^{2i\pi \vec l'\cdot (\vec a_{i_0'}-\vec a_{i'})} .
\end{align}
In the above definitions, ${\rm v}_2\equiv R_4R_5$ and ${\rm v}_4\equiv R_6R_7R_8R_9$, while $\vec a_{i'}, \vec a_i$ are respectively two- and four-vectors parametrizing the positions of the fixed points $i'$ and $i$ in the T-dual pictures: $2\pi R_{I'}a_{i'}^{I'}\in\{0,\pi R_{I'}\}$, $2\pi R_I a_{i}^{I}\in\{0,\pi R_{I}\}$. For the M\"obius strip, the normalization functions are 
\begin{equation}
\begin{aligned}
& \C_{1}^{\M\vec l'\vec l}=0\, ,\qquad&&{\tilde{\C}}_{1}^{\M\vec l'\vec \tl}=0\, , \\
& \C_{2}^{\M\vec l'\vec l}={\C\, \vartheta_2^2\over \tau_2^4\, \eta^9}\, {\rmv_2\rmv_4\over \alpha'^3}\, e^{2i\pi\vec l'\cdot \vec a_{\rm S}'}\, , \qquad&&{\tilde{\C}}_{2}^{\M\vec l'\vec \tl}={\C\, \vartheta_2^2\over \tau_2^4\, \eta^9}\, {\rmv_2\,\alpha'^2\over \alpha'\, \rmv_4}\, e^{2i\pi\vec l'\cdot \vec a_{\rm S}'}\,,\\
& \C_{3}^{\M\vec l'\vec l}=-{\C\, \vartheta_3^2\over \tau_2^4\, \eta^9}\, {\rmv_2\rmv_4\over \alpha'^3}\, ,\qquad&&{\tilde{\C}}_{3}^{\M\vec l'\vec \tl}=-{\C\, \vartheta_3^2\over \tau_2^4\, \eta^9}\, {\rmv_2\,\alpha'^2\over \alpha'\, \rmv_4}\, ,\\
& \C_{4}^{\M\vec l'\vec l}={\C\, \vartheta_4^2\over \tau_2^4\, \eta^9}\, {\rmv_2\rmv_4\over \alpha'^3}\, ,\qquad&&{\tilde{\C}}_{4}^{\M\vec l'\vec \tl}={\C\,\vartheta_4^2\over \tau_2^4\, \eta^9}\, {\rmv_2\,\alpha'^2\over \alpha'\, \rmv_4}\, .
\end{aligned}
\end{equation}

\section{Squared masses in the limit of small \bm $M_{3/2}$}
\label{limits}

In principle, the squared masses of the moduli arising from the ND+DN sector can be extracted from the internal parts of the amplitudes. However, all expressions being so far complicated, it is  illuminating to derive the masses in the regime where the scale of supersymmetry breaking is lower than all other mass scales of the classical spectrum. 

This can be done by considering a field theory limit $\alpha'\to 0$ which implies all string-oscillator states to be supermassive. To this end, the imaginary parts of the Teichm\"uller parameters of the double-covering tori are rescaled,
\begin{equation}
\Im \taudc\equiv {\tau_2\over 2}\equiv {t\over 2\pi \alpha'}\gg 1\, , \quad \where \quad t\in(0,+\infty)\, .
\end{equation}
Moreover, the imaginary parts of the insertion points can be redefined as 
\begin{equation}
\Im z_A= u_A\, \Im\taudc={t\,u_A\over 2\pi\alpha'}\gg 1\,, \quad u_A\in(0,1)\, ,~~A\in\{1,2\}\, .
\end{equation}
For all compactification scales to be greater than $1/R_5$, we also rescale all radii and T-dual radii other than $R_5$ as follows,
\begin{equation}
R_4=r_4\sqrt{\alpha'}\ll 1\, ,\quad R_I=r_I\sqrt{\alpha'}\ll1\, , \quad {\alpha'\over R_I}={\sqrt{\alpha'}\over r_I}\ll 1\, ,\quad \mbox{$r_4$, $r_I$ finite}\, .
\end{equation}
In this regime, the light states of the theory are the Kaluza--Klein modes propagating in the Scherk--Schwarz direction $X^5$. 

Before taking the small $\alpha'$ limit in Eq.~(\ref{Aint}), it is convenient to apply a Poisson summation over the indices $l_4, l_I$ and $\tilde l_I$. The reason is that the dominant contributions arise from zero modes of the new lattices, while all other terms are exponentially suppressed. In the brackets appearing in the expression of $A_{\text{int}\Sigma}^{\alpha_0\beta_0}$, the  terms involving $v_2^u,\tilde v_2^u$ and those proportional to $C+\hat C$ arise respectively from part~$(1)$ and part~$(2)$ of the correlator of excited boundary-changing fields. They both yield contributions of the same order of magnitude in the limit of small~$\alpha'$. Integrating over the proper times $t$ of the diagrams, as well as over the position of one insertion point along one boundary of the annulus and the single boundary of the M\"obius strip, one obtains the squared mass\cite{CP} 
\begin{align}
M^2={32\over \pi}\, \C s \sum_{l_5}{\tr(\lambda\lambda^{\rm T})\over  |2l_5+1|^3} \,M_{3/2}^2\,(n_{i_0i_0'}-n_{i_0\hat \imath_0'}-1+d_{j_0i_0'}-d_{j_0\hat \imath_0'}-1)+\O(\alpha' M_{3/2}^4) \, .
\end{align}
In this formula, $\hat \imath_0'$ denotes the fixed point in the T-dual pictures whose coordinates satisfy $a^4_{\hat \imath_0'}=a^4_{i_0'}$, $a^5_{\hat \imath_0'}=a^5_{i_0'}+\half\mbox{ modulo 1}$. As expected, the final answer in the regime where $M_{3/2}$ is lower than all other mass scales of the spectrum is dominated by the contributions of the infinite towers of Kaluza--Klein states in $4+1$ dimensions running in the loop. 



\begin{thebibliography}{00}    

\bibitem{Hashimoto}
A.~Hashimoto,
``Dynamics of Dirichlet-Neumann open strings on D-branes,''
Nucl. Phys. B \textbf{496} (1997), 243-258
[arXiv:hep-th/9608127 [hep-th]].

\bibitem{Dixon}
L.~J.~Dixon, D.~Friedan, E.~J.~Martinec and S.~H.~Shenker,
``The conformal field theory of orbifolds,''
Nucl.\ Phys.\ B {\bf 282} (1987) 13.

\bibitem{Atick}
J.~J.~Atick, L.~J.~Dixon, P.~A.~Griffin and D.~Nemeschansky,
``Multiloop twist field correlation functions for $\Z_N$ orbifolds,''
Nucl.\ Phys.\ B {\bf 298} (1988) 1.

\bibitem{Burgess1}
C.~P.~Burgess and T.~R.~Morris,
``Open and unoriented strings \`a la Polyakov,''
Nucl. Phys. B \textbf{291} (1987), 256-284.

\bibitem{Burgess2}
C.~P.~Burgess and T.~R.~Morris,
``Open superstrings \`a la Polyakov,''
Nucl. Phys. B \textbf{291} (1987), 285-333.

\bibitem{BianchiSagnotti}
M.~Bianchi and A.~Sagnotti, 
``Twist symmetry and open string Wilson lines,''
Nucl.\ Phys.\ B {\bf  361} (1991) 519.
 
\bibitem{GimonPolchinski}
E.~G.~Gimon and J.~Polchinski,
``Consistency conditions for orientifolds and D-manifolds,''
Phys.\ Rev.\ D {\bf 54} (1996) 1667
[hep-th/9601038].
  
\bibitem{GimonPolchinski2}
M.~Berkooz, R.~G.~Leigh, J.~Polchinski, J.~H.~Schwarz, N.~Seiberg and E.~Witten,
``Anomalies, dualities, and topology of $D=6$ $\N=1$ superstring vacua,''
Nucl.\ Phys.\ B {\bf 475} (1996) 115
[hep-th/9605184].

\bibitem{openSS2}
J.~D.~Blum and K.~R.~Dienes,
``Strong / weak coupling duality relations for nonsupersymmetric string theories,''
Nucl.\ Phys.\ B {\bf 516} (1998) 83
[hep-th/9707160].

\bibitem{openSS3}
I.~Antoniadis, E.~Dudas and A.~Sagnotti,
``Supersymmetry breaking, open strings and M-theory,''
Nucl.\ Phys.\ B {\bf 544} (1999) 469
[hep-th/9807011].

\bibitem{SS1}
J.~Scherk and J.~H.~Schwarz,
``Spontaneous breaking of supersymmetry through dimensional reduction,''
Phys.\ Lett.\ B {\bf 82} (1979) 60.

\bibitem{ACP}
S.~Abel, T.~Coudarchet and H.~Partouche,
``On the stability of open-string orbifold models with broken supersymmetry,''
Nucl. Phys. B \textbf{957} (2020), 115100
[arXiv:2003.02545 [hep-th]].

\bibitem{CP}
T.~Coudarchet and H.~Partouche,
``One-loop masses of Neumann-Dirichlet open strings and boundary-changing vertex operators,''
arXiv:2011.13725 [hep-th].

\bibitem{Coudarchet:2020ozn}
T.~Coudarchet and H.~Partouche,
``Moduli stability in type~I string orbifold models,''
PoS \textbf{CORFU2019} (2020), 164
[arXiv:2005.01764 [hep-th]].

\bibitem{Abel:2015oxa} 
S.~Abel, K.~R.~Dienes and E.~Mavroudi,
``Towards a non-supersymmetric string phenomenology,''
Phys.\ Rev.\ D {\bf 91} (2015) 126014
[arXiv:1502.03087 [hep-th]].
  
\bibitem{SNS1}
C.~Kounnas and H.~Partouche,
``Super no-scale models in string theory,''
Nucl.\ Phys.\ B {\bf 913} (2016) 593
[arXiv:1607.01767 [hep-th]].
  
\bibitem{review-3}
J.~Polchinski,
``Tasi lectures on D-branes,''
hep-th/9611050.

\bibitem{Abel}
S.~A.~Abel and B.~W.~Schofield,
``One-loop Yukawas on intersecting branes,''
JHEP \textbf{06} (2005), 072
[arXiv:hep-th/0412206 [hep-th]].

\end{thebibliography}
\end{document}